\documentclass[12pt,a4paper]{article}
\usepackage[T2A]{fontenc}
\usepackage[cp1251]{inputenc}
\usepackage[english]{babel}

\usepackage{booktabs}
\usepackage{varwidth}
\usepackage{ulem,xcolor}

\usepackage{setspace}

\usepackage{amsmath,amsfonts,stackengine}
\usepackage{bm,latexsym,euscript,textcomp}
\usepackage{graphicx,color,graphics,epsf,dcolumn}
\usepackage{natbib}
\usepackage{epstopdf}

\newcommand{\beq}{\begin{equation}}
\newcommand{\eeq}{\end{equation}}
\newcommand{\pt}{\partial}

\begin{document}

\title{\Large \bf  Hurricane's Maximum Potential Intensity and the Gravitational Power of Precipitation}

\author{A. M. Makarieva$^{1,2}$\thanks{\textit{Corresponding author.} {E-mail: ammakarieva@gmail.com}}, V. G. Gorshkov$^{1,2}$, A.V. Nefiodov$^1$, A.V. Chikunov$^{3,4}$,  \\ D.  Sheil$^{5}$,  A. D. Nobre$^6$, B.-L. Li$^2$}

\date{\vspace{-5ex}}

\maketitle

\noindent
$^1$Theoretical Physics Division, Petersburg Nuclear Physics Institute, 188300 Gatchina, St. Petersburg, Russia
$^2$USDA-China MOST Joint Research Center for AgroEcology and Sustainability, University of California, Riverside 92521-0124, USA
$^3$Institute of World Ideas, Udaltsova street 1A, Moscow 119415, Russia
$^4$Princeton Institute of Life Sciences, Princeton, USA
$^5$Faculty of Environmental Sciences and Natural Resource Management, Norwegian University of Life Sciences, \AA s, Norway
$^6$Centro de Ci\^{e}ncia do Sistema Terrestre INPE, S\~{a}o Jos\'{e} dos Campos, S\~{a}o Paulo  12227-010, Brazil

\begin{abstract}
The concept of Maximum Potential Intensity (MPI) is widely used in tropical cyclone research to estimate the minimum central pressure and the maximum velocity of tropical storms from environmental parameters. The MPI pressure derives from consideration of an idealized thermodynamic cycle, while the MPI velocity additionally requires information about real-time power and heat flows within the storm. Recently MPI velocity was proposed to be a substantial overestimate (by 10-30 percent) presumably neglecting the power needed to lift precipitating water (the gravitational power of precipitation). This conclusion did not involve a theoretical analysis of the MPI concept but was based on observed hurricane rainfall to estimate  gravitational power of precipitation. However, since the MPI pressure estimate does explicitly account for lifting water, and the MPI velocity derives from this pressure, the question arises whether a correction to MPI velocity is required. Here we represent and examine the MPI derivations in their most general form and show that although a correction to MPI velocity is justified, it is an order of magnitude or so smaller than originally proposed. We show that the neglect of  gravitational power of precipitation  in the MPI velocity estimate  was caused by the incomplete formulation of the general relationship between pressure work and dissipation of kinetic energy, taken per unit time and integrated over the storm. We highlight the importance of an internally consistent framework to estimate both storm energy and power and provide some perspectives for further investigating the role of moisture.
\end{abstract}

\section{Introduction} 

\label{int}

Predicting hurricane intensity is a challenge in atmospheric research. Historically, much theoretical attention focused on finding upper limits on  hurricane intensity \citep{malkus60,holland97,CampMontgomery01}. While {\it intensity} conventionally denotes maximum sustained velocity within a storm, early theoretical studies rather sought to estimate minimum central pressure known to be well correlated with maximum velocity.

Given that the hurricane is warmer than the ambient environment the idea was to retrieve the surface pressure deficit from this extra warmth assuming the existence of an unperturbed atmospheric top where pressures in the hurricane and the environment coincide. Since air pressure drops with altitude more slowly when the atmosphere is warm than when it is cold, to arrive at equal pressures at the top of the troposphere one must start from a lower surface pressure in the warmer column. The height of the unperturbed top and the extra warmth of the hurricane compared to its environment uniquely determined the surface pressure deficit in the storm.

\citet{em86} advanced beyond this static approach by noting that pressure work (which produces the kinetic energy of wind)
is constrained not only by the first law of thermodynamics, but also by the Bernoulli equation that derives from the equations of motion and continuity. Combining these two perspectives and additionally assuming that both generation of the kinetic energy and its dissipation (proportional to the cube of velocity) occur within the boundary layer, \citet{em86} thus linked {\it work} to {\it power} to
estimate maximum velocity for a given central pressure. With its explicit formulae for calculating maximum hurricane velocity
from environmental parameters, this approach became widely used in the tropical storm community.

\citet{em88,em91,em95,em97} further advanced the MPI concept through several major modifications aimed to remedy the limitations of the original formulation. It was not before late 2000s that several evaluations of the concept were provided by independent theorists \citep{sm08,dhe10,smith14,kieu15}. Those studies  did not consider the MPI concept in its integrity, each focusing either on its dynamic or thermodynamic aspects. Recently, \citet{sabuwala15} proposed that the MPI velocity is a substantial overestimate since it presumably does not account for lifting water vapor.

Here it is essential to note that the MPI concept comprises two distinct approaches to storm's energetics (Fig.~\ref{fig1}).
One considers {\it work} in a steady-state thermodynamic cycle and kinetic energy variation along a closed streamline
corresponding to that cycle (units J~kg$^{-1}$) (Fig.~\ref{fig2}).  Another considers real-time heat flows and wind {\it power} (work per unit time) within the storm (units J~s$^{-1}$).  \citet{em86} calculated the minimum central pressure in the storm using the {\it work} approach; then used the obtained value in the {\it power} approach to calculate storm's maximum velocity. Power cannot be estimated from work without adding the scale of time. Furthermore, proceeding from an idealized Carnot cycle to real-time heat engines requires introducing
additional parameters describing the steady-state disequilibrium between the heat source and the working body \citep[e.g.,][]{curzon75}.
At the same time the work and power approaches are obviously interdependent and should be combined coherently.

Here we analyze the MPI concept from this perspective and show that while in the {\it work} approach the energy needed for lifting water has been indeed accounted for by \citet{em88}, this has not been done in the {\it power} approach (Fig.~\ref{fig1}). We show that in the presence of phase transitions the relationship between pressure work and kinetic energy dissipation is not the same in the work and power approaches and that the neglect of this distinction is responsible for the omission of the gravitational power of precipitation in the MPI velocity estimate.

\section{Deriving MPI}

\label{DerMPI}

The MPI concept views the hurricane as a thermodynamic cycle consuming heat from the ocean (Fig.~\ref{fig2}). Work performed
in the cycle per unit mass of dry air is
\beq\label{wc}
-\oint \alpha_d dp = \varepsilon Q,
\eeq
where $\alpha_d \equiv 1/\rho_d$ is the specific volume of dry air, $\rho_d$ is dry air density, $Q$ (J~kg$^{-1}$) is
heat input from the ocean per unit dry air mass and $\varepsilon$ is the cycle's efficiency of converting heat to work.

Additionally, the MPI concept employs the Bernoulli equation
\beq\label{B}
d\left(\frac{v^2}{2}\right) + \alpha dp + g dz - \mathbf{f} \cdot d\mathbf{l} = 0,
\eeq
where $v$ is air velocity, $\alpha \equiv 1/\rho$, $\rho$ is the density of moist air,
$\mathbf{f}$ is the friction force per unit air mass and $d\mathbf{l}=\mathbf{v}dt$,
see \citet[][Eq.~64]{em86} and \citet[][Eq.~C1]{em88}. From Eq.~(\ref{B}) we have
\beq\label{f}
-\oint \alpha dp = -\oint \mathbf{f} \cdot d\mathbf{l},
\eeq
which can be interpreted as a "balance between pressure work and dissipation in steady flow" \citep{em88}.

To proceed from work to real-time power, one has to replace $dp$ in Eqs.~(\ref{wc}) and (\ref{f})
by $dp/dt$, where $d/dt$ is material derivative, and integrate them over the entire hurricane occupying volume $\mathcal{V}$
with total air mass $\mathcal{M}$ and total dry air mass $\mathcal{M}_d$.
Taking into account that $d\mathcal{V} = d\mathcal{M}_d/\rho_d = d\mathcal{M}/\rho$ we find using Eq.~(\ref{wc}) (cf. Eqs.~(w1)
and (p1) in Fig.~\ref{fig1})
\beq\label{int}
-\int\limits_{\mathcal{M}_d} \alpha_d \frac{dp}{dt} d\mathcal{M}_d = -\int\limits_{\mathcal{M}} \alpha \frac{dp}{dt} d\mathcal{M} =
-\int\limits_{\mathcal{V}} \frac{dp}{dt} d\mathcal{V} = \varepsilon J,
\,\,\, J \equiv \int_{z \le h_b} \dot{Q} d\mathcal{M}_d.
\eeq
Here $J$ (W) is the total heat flow from the ocean into the hurricane,
$\dot{Q}$ (W~kg$^{-1}$) is the local heat source per unit mass of dry air,
$h_b$ is the height of the boundary layer beneath which this heat intake from
the oceanic surface is assumed to occur (Fig.~\ref{fig2}).

On the other hand, integrating the Bernoulli equation over $\mathcal{M}$
using the definition of material derivative
\beq\label{der}
\frac{dp}{dt} = \mathbf{v} \cdot \nabla p,
\eeq
and the continuity equation for gaseous air
\beq\label{cont}
\nabla \cdot (\rho \mathbf{v}) = \dot{\rho},
\eeq
where $\dot{\rho}$ (kg~s$^{-1}$~m$^{-3}$) is the volume-specific rate of phase transitions,
we find
\beq\label{bud}
-\int\limits_{\mathcal{V}} \frac{dp}{dt} d\mathcal{V}= -\int\limits_{\mathcal{V}} \mathbf{F}\cdot \mathbf{v} d\mathcal{V} - \int\limits_{\mathcal{V}} \dot{\rho} g z d\mathcal{V},
\eeq
see Appendix for details. Here $\mathbf{v}$ is the velocity vector for gaseous air, $\mathbf{F} \equiv \rho \mathbf{f}$ is
friction force per unit air volume.

The second term in the right-hand part of Eq.~(\ref{bud}) has the meaning of the gravitational power of precipitation.
Since condensation predominantly occurs for $z > 0$ (and evaporation is concentrated at the surface $z = 0$),
we have on average $\dot{\rho} < 0$ for $z > 0$. With $gz$ being potential energy per unit mass, this term
is positive and represents the rate at which condensation creates condensate with potential energy $gz$.
For the real-time energy flows in the atmosphere pressure work per unit time given by the first term in Eq.~(\ref{bud}) is {\it not} balanced
by production/dissipation rate of kinetic energy alone but by the production/dissipation rate of kinetic energy {\it and} the gravitational power of precipitation.

To our knowledge, equation~(\ref{bud}) and its derivation (Appendix) have not been previously described.
While \citet{pa00} proposed that {\it total atmospheric power} should be equal to the sum of kinetic energy production/dissipation and the gravitational power of precipitation, they did not show that {\it total atmospheric power} is equal to $-\int_\mathcal{V} (dp/dt) d\mathcal{V}$. Thus, the idea that pressure work is balanced by kinetic energy dissipation as per Eq.~(\ref{f}) and the common use of continuity equations with a zero source term $\dot{\rho}=0$ caused a wide-spread misinterpretation of $-\int_\mathcal{V} (dp/dt) d\mathcal{V}$ for the rate of kinetic energy production and dissipation \citep[e.g.,][]{pa15}. This led to an overestimation of the rate of kinetic energy production in the MPI concept as discussed below.

\section{MPI velocity estimate}

Combining Eqs.~(\ref{int}) and (\ref{bud}) we find
\beq\label{eps}
W_K = \left(\varepsilon - \frac{W_P}{J}\right) J,
\eeq
where $W_K$ (W) is the rate of kinetic energy production and dissipation
and $W_P$ (W) is the gravitational power of precipitation:
\beq\label{def}
W_K \equiv -\int\limits_{\mathcal{V}} \mathbf{F}\cdot \mathbf{v} d\mathcal{V},\,\,\, W_P \equiv -\int\limits_{\mathcal{V}} \dot{\rho} g z d\mathcal{V} > 0.
\eeq

Equations (\ref{wc})-(\ref{def}) that we have so far considered are generally valid. Specific to the MPI concept is
the assumption that dissipation of kinetic energy predominantly occurs in the boundary layer, such that
\beq\label{dis1}
\oint \mathbf{f}\cdot d\mathbf{l} = \int_a^c \mathbf{f}\cdot d\mathbf{l}
\eeq
in energy units  (see, e.g., Eq.~C4 of \citet{em88} where the last minor term is commonly neglected)
and
\beq\label{dis2}
W_K \equiv -\int\limits_{\mathcal{M}} \mathbf{f}\cdot \frac{d \mathbf{l}}{dt} d\mathcal{M} = -\int\limits_{z \le h_b} \mathbf{F}\cdot \mathbf{v} d\mathcal{V} =
\int\limits_{\mathcal{S}} \rho C_D v^3 d\mathcal{S}
\eeq
in power units \citep[e.g., Eq.~7 of~][]{em97}. Expression under the last integral in Eq.~(\ref{dis2}) is the local dissipation rate
in the boundary layer per unit surface area, $C_D \sim 10^{-3}$ is a dimensionless coefficient,
the integration is made over area $\mathcal{S}$ occupied by the hurricane.

For the heat input from the ocean the MPI concept uses the following relationship \citep[see, e.g.,][Eq.~3]{em95}:
\beq\label{heat}
J = \int\limits_{\mathcal{S}} \rho C_k v(k_s^* - k) d\mathcal{S},
\eeq
where $C_k \sim 10^{-3}$, $k_s^*$ (J~kg$^{-1}$) is saturated enthalpy of air at surface temperature
and $k$ is the actual enthalpy of air in the boundary layer.

Combining Eqs.~(\ref{eps}), (\ref{dis2}) and (\ref{heat}), we find
\beq\label{res}
\int\limits_{\mathcal{S}} \rho C_D v^3 d\mathcal{S} = \varepsilon_K \int\limits_{\mathcal{S}} \rho C_k v (k_s^* - k) d\mathcal{S},\,\,\, \varepsilon_K \equiv \varepsilon - \frac{W_P}{J} < \varepsilon.
\eeq

A major assumption within the MPI concept is that Eq.~(\ref{res}) (but with $\varepsilon_K$ replaced by
$\varepsilon$) holds for the expressions under the integral in the region of maximum velocities \citep{em97}, such that
from Eq.~(\ref{res}) we would have
\beq\label{vmax}
v_{\rm max}^2 = \varepsilon_K \frac{C_k}{C_D} (k_s^* - k).
\eeq
We discuss this assumption below. Here we note that $\varepsilon$ and $\varepsilon_K$
are not local characteristics but those of the thermodynamic {\it cycle} and the considered closed air trajectory {\it as a whole}.
Therefore, if the major inputs into Eq.~(\ref{res}) are indeed made in the region of $v = v_{\rm max}$, as assumed in the MPI concept,
the reduction of efficiency and replacement of $\varepsilon$ by $\varepsilon_K$ in Eq.~(\ref{res}) will also apply
to the velocity estimate $v_{\rm max}$ (\ref{vmax}).

We can represent the gravitational power of precipitation as
\beq\label{WP}
W_P = Pg H_P,
\eeq
where $H_P$ is the mean height at which condensation occurs \citep{vg80,pa11,jas13} and $P$ (kg~s$^{-1}$) is total precipitation
over area $\mathcal{S}$ of the storm.

For the oceanic heat input we have
\beq\label{Q}
J = J_S + J_L = (B+1) PL,
\eeq
where $J_S$ and $J_L$ are fluxes of sensible and latent heat, respectively; $B \equiv J_S/J_L$ is the Bowen ratio,
$L =2.5\times 10^6$~J~kg$^{-1}$ is the latent heat of vaporization and $P$ is rainfall here assumed to be equal to the flux
of evaporation to be consistent with the steady-state MPI concept.

From Eqs.~(\ref{res}), (\ref{WP}) and (\ref{Q}) we find
\beq\label{epsK}
\varepsilon_K = \varepsilon  - \frac{g H_P}{(B+1) L}.
\eeq

Mean precipitation height $H_P$ can be calculated from the equation of moist adiabat and depends on surface temperature $T_s$,
the incompleteness of condensation $\zeta$ and, to a lesser degree, on surface relative humidity \citep{jas13}.
We assume that moist air having temperature $T_s$
and relative humidity $80\%$ at the surface first rises dry adiabatically
up to height $z_1$ where water vapor becomes saturated. Then it rises moist adiabatically
to $z_2$, where condensation ceases. At $z_2$
the air preserves share $\zeta$ of its initial water vapor content, $\zeta \equiv \gamma(z_2)/\gamma_s = \gamma(z_2)/\gamma(z_1)$.
Here $\gamma \equiv p_v/p$, where $p_v$ is water vapor partial pressure and $p$ is air pressure.
Moist adiabatic distributions of $\gamma(z)$ and $p(z)$ with $p_s = 1000$~hPa,
where subscript $s$ denotes surface values, were calculated according to Eqs.~(A3)-(A5) of \citet{jas13}

For $T_s$ ranging from $260$ to $310$~K and for $\zeta$ ranging from $0.001$ (almost complete condensation)
to $1/2$, we estimated mean condensation height as
\beq\label{HP}
H_P(T_s,\zeta) = \frac{1}{\gamma(z_2)-\gamma(z_1)}\int_{z_1}^{z_2} z \frac{\pt \gamma}{\pt z} dz.
\eeq

Height $H_P$ grows with increasing temperature but even for complete removal of
water vapor from the air, which always occurs if the air rises above 16 km, as is often the case in hurricanes,
$H_P$ does not exceed 6~km (Fig.~\ref{fig3}).

For a typical Bowen ratio of 1/3 \citep{jaimes15} with
$H_P = 6$~km we obtain from Eq.~(\ref{epsK}) that the gravitational power of precipitation reduces $\varepsilon_K$ compared to $\varepsilon$ by $gH_P/L \sim 0.018$ at most. With $\varepsilon$ for
hurricanes viewed as Carnot cycles being around $0.3$ \citep{em86,demaria94}, this represents a 6\% reduction to
$\varepsilon$ and a 3\% reduction to velocity $v_{\rm max}$ (\ref{vmax}). For circulations with a smaller $\varepsilon$
the relative reduction would be larger.

\citet{sabuwala15} investigated how accounting for the gravitational power of precipitation can reduce maximum velocity compared to its MPI estimate. They did not present a theoretical estimate of $\varepsilon_K$ but used instead empirical TRMM data on hurricane rainfall in the vicinity of maximum velocity. Their conclusion was a 10-30\% reduction in velocity, which
is significantly higher than our estimate of $3\%$.

The reason for those overly high figures is twofold. First, \citet{sabuwala15} used local values of rainfall measured
in the vicinity of maximum velocity, while the reduction pertains to the considered
cycle as a whole and thus should be estimated using mean rainfall within the storm. Mean rainfall
within the outermost closed isobar as estimated from TRMM data is several times lower
than maximum rainfall in the vicinity of the radius of maximum winds. For North Atlantic hurricanes
it is 2~mm~hr$^{-1}$ within 400 km and about 8~mm~hr$^{-1}$ within 100 km which includes the radius of maximum wind \citep[see, e.g., Fig 4j of][]{ar17}.

Second, of this rainfall only about one quarter or third is represented by moisture evaporated within
the outermost closed isobar. The major part of precipitating water is imported from outside \citep{ar17}.
The MPI concept does not explicitly account for this imported moisture as it views the
hurricane as a steady-state thermodynamic cycle with moisture provided locally by evaporation from the ocean.
This imported moisture is, however, implicitly accounted for by considering the hypothetical adiabat $o'-a$
to be part of the hurricane's thermodynamic cycle (Fig.~\ref{fig2}).

Along $o'-a$ the moisture content of the hypothetically descending air rises by over two orders of magnitude.
In this region, i.e. outside the radius of the outermost closed isobar $r_o$ (Fig.~\ref{fig2}), there
is a characteristic {\it clear-sky moat} \citep{frank77,ar17}. Thus this increase of moisture content
cannot occur due to "mixing with cloudy air" \citep[cf.][]{pa17}.
Indeed, most moisture condensing within the hurricane precipitates and cannot serve as a source of water vapor
for the descending air. The vertical distribution of humidity along the $o'-a$ path is in fact provided by evaporation and convection outside the storm. These moisture stores are picked up by the hurricane as it moves through the atmosphere. As the moisture is imported with its own gravitational energy, the storm does not need to spend energy and power to raise this water.

\begin{figure*}[!htb]
\vspace{-0.5 cm}
\begin{minipage}[p]{\textwidth}
\centering\includegraphics[width=0.8\textwidth,angle=0,clip]{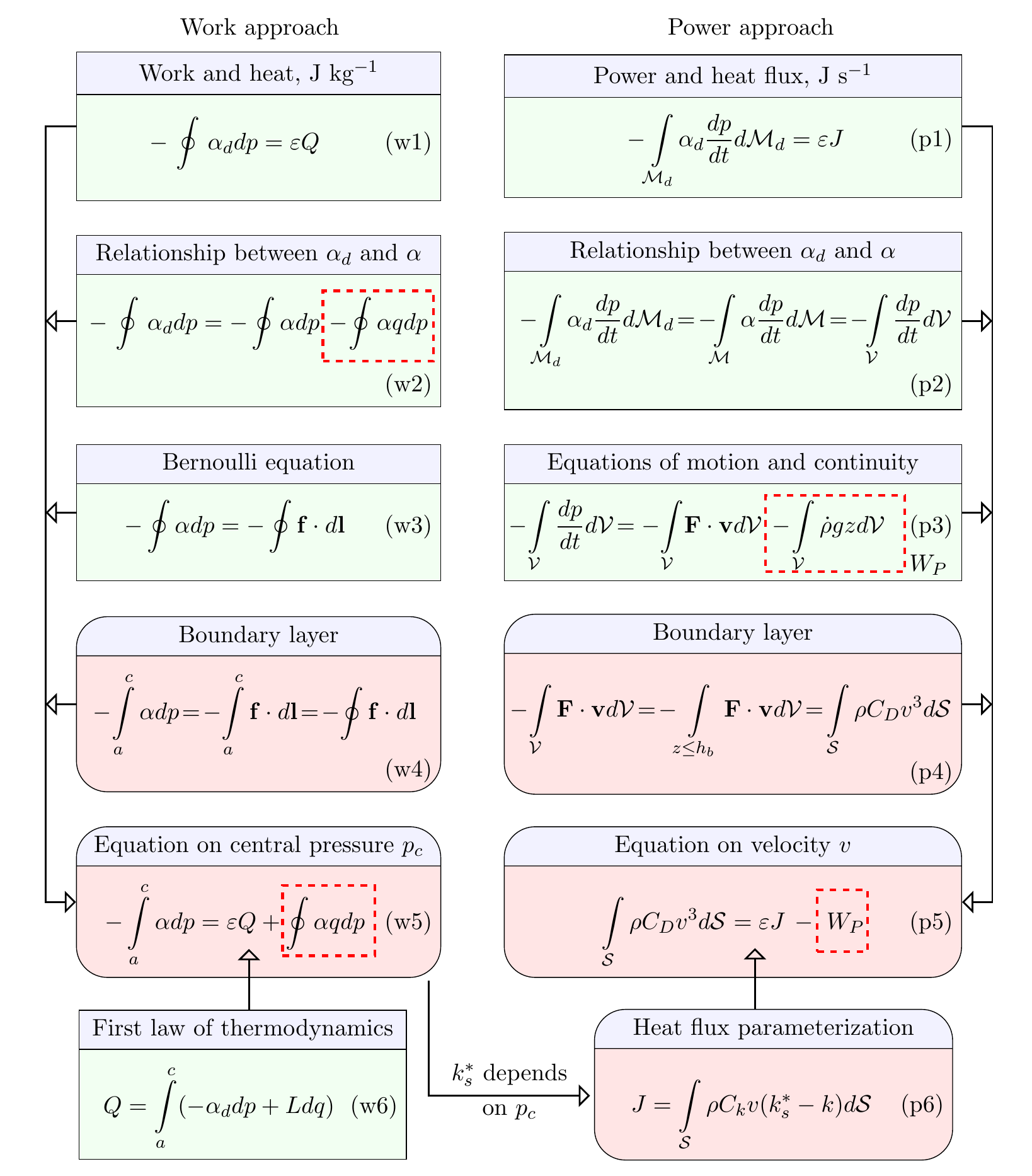}
\end{minipage}
\caption{
Logical structure of the MPI concept. Rectangular boxes enclose generally valid statements, rounded boxes enclose assumptions specific to the MPI concept, $q \equiv \rho_v/\rho_d$, see text for the other notations. Dashed frames enclose terms accounting for the energy (left column) and power (right column) needed to lift water, $W_P$ is the gravitational power of precipitation. Relationships in the left column are those given by \citet{em88}. In the original paper of \citet{em86} the difference between $\alpha$ and $\alpha_d$ and, hence, the energy needed to lift water was neglected. Term $W_P$ in Eqs.~(p3) and (p5) was not accounted for in the MPI derivations.}
\label{fig1}
\end{figure*}

\begin{figure*}[!htb]
\vspace{-0.5 cm}
\begin{minipage}[p]{\textwidth}
\centering\includegraphics[width=0.8\textwidth,angle=0,clip]{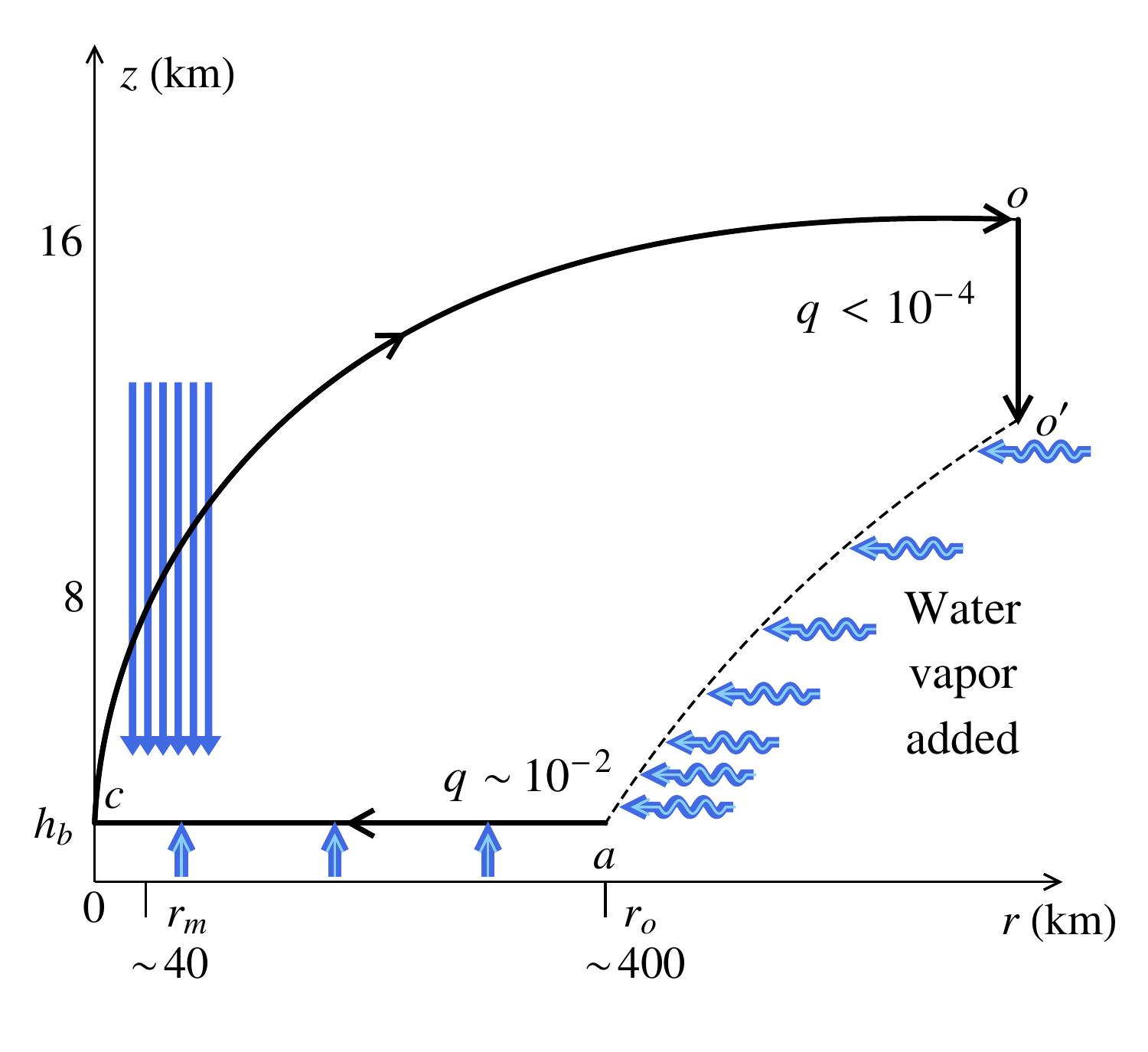}
\end{minipage}
\caption{A hurricane in the MPI concept. Solid curve $a-c-o-o'$ is the actual streamline of air entering the hurricane at point $a$ and leaving it at point $o'$; dashed curve $o'-a$ is the hypothetical path closing the thermodynamic cycle $a-c-o-o'-a$ with two isotherms, $a-c$ and $o-o'$, and two adiabats, $c-o$ and $o'-a$. As the air rises from $c$ to $o$, water vapor condenses and precipitates, water vapor mixing ratio $q \equiv \rho_v/\rho_d$, where $\rho_v$ is water vapor density, $\rho_d$ is dry air density, declines by over two orders of magnitude.
A major part of this lost water re-appears in the cycle (as shown by waved arrows) along the hypothetical $o'-a$ adiabat; in the real hurricane this imported moisture derives from evaporation outside the storm and is picked up as the storm moves through the atmosphere.
The remaining part of moisture lost as rainfall is provided by evaporation from the sea surface (straight upward arrows).
Straight downward arrows indicate the rainfall maximum that occurs in the vicinity of the radius of maximum wind $r=r_m$; $r_o$ corresponding to point $a$ is an external radius of the storm estimated to be approximately an order of magnitude larger than $r_m$ 
\citep[][Table~1]{em95}; $z = h_b$ is the height of the boundary layer.}
\label{fig2}
\end{figure*}

\begin{figure*}[!htb]
\vspace{-0.5 cm}
\begin{minipage}[p]{\textwidth}
\centering\includegraphics[width=0.8\textwidth,angle=0,clip]{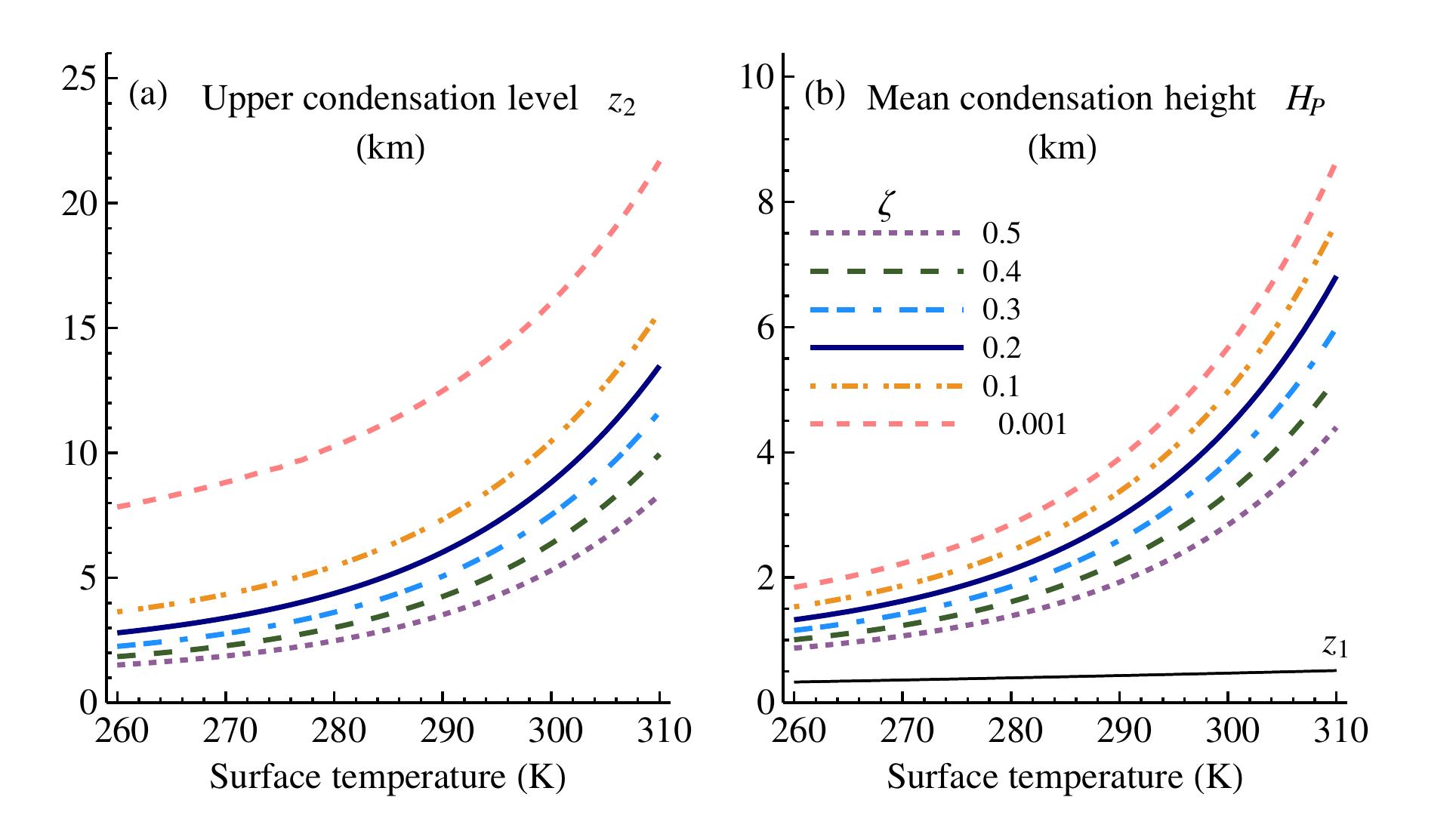}
\end{minipage}
\caption{The upper condensation level $z_2$ (a), lower condensation level $z_1$ and mean condensation height $H_P$ (\ref{HP}) (b),
as dependent on surface temperature $T_s$ and incompleteness of condensation $\zeta$ calculated following \citet{jas13}. }
\label{fig3}
\end{figure*}

This issue was addressed by \citet{em88} within the work approach (see the left column in Fig.~\ref{fig1}).
Comparing Eqs.~(\ref{f}) and (\ref{bud}) (cf. also Eqs.~(w3) and (p3) in Fig.~\ref{fig1}) we can see that while the integrals of $\alpha dp$ and $\mathbf{f}\cdot d\mathbf{l}$ (J~kg$^{-1}$) over closed streamlines coincide, the integrals of $\alpha dp/dt$ and $\mathbf{f}\cdot d\mathbf{l}/dt$ (W~kg$^{-1}$) over the hurricane mass are not equal. The reason is that integration over the closed streamline is made for a constant unit mass. Meanwhile in the real-time atmosphere in the presence of phase transitions there is simultaneously more gas rising and expanding (positive work) than descending and compressing (negative work). The difference between these amounts of gas is what accounts for the gravitational power of precipitation.

Accordingly, \citet{em88} showed that work of the thermodynamic cycle is equal to the sum of kinetic energy generation and the net work of rising (expanding) versus compressing (descending) water vapor expressed by the term $\oint \alpha q dp$, see Eqs.~(w2) and (w5) in Fig.~\ref{fig1}.  \citet{em88} indicated that the term accounting for water lifting energy is proportional to the difference in the water profiles of the air rising along $c-o$ path and the environmental air along the hypothetical path $o'-a$ \citep[][see Appendix C and Eq.~(C12)]{em88}. Recently \citet{pa17} repeated these derivations to retrieve the energy needed to lift water from a numerical hurricane model. Neither \citet{em88} nor \citet{pa17} estimated the contribution of $W_P$ theoretically (this requires an estimate of $H_P$ as per Fig.~\ref{fig3}).

Thus, within the work approach which allows the estimation of hurricane's pressure profile and its minimum central pressure $p_c$ (which in turn impacts the value of saturated enthalpy $k_s^*$ in the expression for $v_{\rm max}$, see Eq.~(w5) in Fig.~\ref{fig1}) the energy needed to lift water was accounted for by \citet{em88}. This requires recognizing the difference between $\alpha$ and $\alpha_d$ (Fig.~\ref{fig1}) which was not accounted for in the original paper by \citet{em86} and in subsequent papers \citep[e.g.,][]{em91}. Indeed, while local values of $\alpha_d$ and $\alpha$ are very close (the difference is of the order of water vapor mixing ratio $q \sim 10^{-2}$), their integrals over the closed cycle differ by an amount comparable to the two integrals themselves: all terms in Eq.~(w2) in Fig.~\ref{fig1} can be of the same order of magnitude. In the power approach the gravitational power of precipitation has not been so far accounted for.

Returning to the estimates of \citet{sabuwala15}, overestimating the surface-specific rainfall in (\ref{WP})
is equivalent to overestimating $H_P$ in (\ref{epsK}) by the same factor. Using local rainfall instead of mean hurricane
rainfall (the overestimate factor of about 4) and all rainfall instead of rainfall provided by evaporation from the hurricane
(the overestimate factor of 3-4), they should have overestimated the actual correction to $v_{\rm max}$ by
an order of magnitude. This explains the difference between their maximum proposed reduction of $30\%$ and our estimate
of $3\%$.

\section{Discussion}

We have shown that the reduction in MPI velocity estimate associated with the gravitational power of precipitation
was previously overestimated but that this can be addressed using Eq.~(\ref{bud}). This equation shows that, unlike in the work approach where the work due to pressure is balanced by frictional dissipation (Eq.~(w3) in Fig.~\ref{fig1}), in the power approach work per unit time of the pressure gradient forces is balanced by the rate of frictional dissipation of kinetic energy plus the gravitational power of precipitation (Eq.~(p3) in Fig.~\ref{fig1}).

Given the prominence of the MPI concept in the tropical storms research we believe that it would be useful to re-visit the other assumptions within the concept in the view of the constraints imposed by a simultaneous consideration of Eqs.~(w1)-(p6) in Fig.~\ref{fig1}. Below we outline several perspectives for such an analysis.

First, the MPI concept is based on a relationship between angular momentum and moist entropy used
to justify the transition from the integral equation (\ref{dis2}) to the local equation (\ref{vmax}).
This logic is based on the following assumptions: 1) that air at $z = h_b$ (height of the boundary layer) is saturated;
2) that it is isothermal and 3) that air at $z = 0$ is also isothermal \citep[for details see][in particular, the first two unnumbered
equations on page 589, right column]{em86}. At the same time, Eq.~(w6) in Fig.~\ref{fig1} shows that the air moving from $a$ to $c$ increases its water vapor content, i.e. $q$ grows from $a$ to $c$. It is easy to see that these four conditions,
air saturated and isothermal at $z=h_b$, air isothermal at $z=0$ and $q$ at $z=h_b$ growing from $a$ to $c$ are not compatible
with each other. (In brief, for the isothermy at both $z=0$ and $z=h_b$ to be satisfied, the lapse rate below $z=h_b$ must be the
same everywhere; i.e. it must be dry adiabatic (since at different $q$ the moist adiabatic lapse rate will not be the same).
However, saturation height $z_1$ for a given temperature depends on $q$: the higher the $q$, the lower the saturation height;
when $q$ is saturated, $z_1= 0$. So, if $q$ increases from $a$ to $c$, the level at which it becomes saturated
diminishes. Thus, if at point $c$ the air is saturated at $z=h_b$ but not below, it cannot be saturated at point $a$ (with a smaller $q$) at the same  height $z=h_b$, thus the first condition is violated.)

Second, since the MPI concept assumes that the surface air is isothermal, and there is no other reason for this isothermy rather than the isothermal oceanic surface, the concept presumes that the temperature of the surface air and the ocean coincide \citep[see, e.g.,][Table~1, third row, forth column]{holland97}. This means that the flux of sensible heat $J_S$ from the ocean to the hurricane is absent. All energy input consists in the flux of latent heat $J_L$. However, the first law of thermodynamics (w6) in Fig.~\ref{fig1} shows that total heat $Q$ input into the air as it moves from $a$ to $c$ is not confined to latent heat alone $Q_L = \int_a^c Ldq$. It also includes the term $\alpha_d dp$:
the air expands but remains isothermal gaining heat from somewhere beyond latent heat input. Thus, while the MPI assumption that the temperature of surface air coincides with that of the ocean prescribes that sensible heat input is zero, the latent heat alone cannot account for the observed isothermal expansion of air parcels. To our knowledge, this problem was not explicitly acknowledged, but \citet{bister98} suggested that there is an extra source of heat: it is the dissipation of kinetic energy generated within the hurricane \citep[see discussion by][]{dhe10,bister11,kieu15}.

\citet{bister98} formulated this extra heat source within the {\it power} approach.
Considering it within the {\it work} approach reveals the following issue. Generation of kinetic energy, $\alpha dp$
and sensible heat input $\alpha_d dp$ practically concide at the isotherm $a-c$ because of the smallness of $q$.
If at some point $x$ we have $-\int_a^x \alpha_d dp = -\int_a^x \alpha dp = -\int_a^x \mathbf{f} \cdot d\mathbf{l}$, this means,
according to the Bernoulli equation (\ref{B}), that $\int_a^x dv^2 = 0$. In other words, if the energy generated
by pressure work $-\alpha dp$ is dissipated by friction $-\mathbf{f}\cdot d\mathbf{l}$
to account for the missing sensible heat $\alpha_d dp$, air velocity in the hurricane cannot
rise as the air moves from $a$ towards the center, which contradicts the observations.
This emphasizes the need to jointly consider both work and power approaches when formulating MPI \citep{prep18}.

Third, we believe that the MPI concept may have a broader dynamic interpretation not confined to the Carnot cycle or a steady-state
case it has so far been linked to.  As the air leaves the boundary layer, it must have sufficient energy to flow away from the hurricane.
This energy can be provided by a pressure gradient in the upper atmosphere \citep{tellus17}:  if the air pressure
in the column above the area of maximum wind is higher than in the ambient environment, it will accelerate
the air outward. However, a significant pressure deficit at the surface precludes the formation of a significant
pressure surplus aloft. On the other hand, namely this pressure deficit is what accelerates the hurricane air in the boundary layer.
Given that the air as it leaves the boundary layer possesses a certain kinetic energy, we can require
that this energy is spent to overcome the negative pressure gradient in the upper atmosphere.
Thus, the air will be flowing outward not at the expense of a pressure gradient but against it,
at the expense of accumulated kinetic energy (propelled by the centrifugal force). Maximum intensity in this case will
be determined from the condition that the kinetic energy accumulated at the expense of pressure deficit
at the surface is enough to overcome the pressure  surplus  in the upper atmosphere for the air to flow away. This condition
should be a valid upper limit to hurricane intensity for any type of air interaction with the ocean, steady and non-steady
circulations alike. As we discuss elsewhere, the MPI concept in its dynamic part expresses these ideas and
thus, in a properly modified form, should have a broader generality than so far assumed \citep{prep18}.

Forth, the hurricane can be divided into two zones. One is $r \ge r_m$ where the air moves across the isobars
and the kinetic power is generated. The other is the eye, which can be approximated by solid body rotation \citep[e.g.,][]{em97}
with a non-zero energy store but zero power. The pressure drop observed across the hurricane likewise consists of two
parts: the pressure drop within the eye represents a {\it store of potential energy} equal to the kinetic energy
of the eye; no power is generated here. The pressure drop from the ambient environment to the
outer border of maximum wind, which is less than a half of the total pressure drop, is what actually generates
hurricane power (of which a minor part goes to maintain the slowly dissipating energy store within the eye). Recognition of this
spatial division into the {\it energy} and {\it power} zones is essential for understanding hurricane's energetics.

Fifth, and arguably most important, \citet{sabuwala15} demonstrated using empirical data that hurricane intensity
is correlated with rainfall intensity. Furthermore, there is a growing
body of evidence revealing correlation between external moisture supply and hurricane power
\citep{krishnamurti93,krishnamurti98,fritz14,ermakov14,ermakov15,fujiwara17}.
However, a qualitative let alone quantitative explanation of these patterns remains elusive.
\citet{sabuwala15} hypothesized that it is a higher intensity
of latent heat release that is associated with a more powerful hurricane. The same logic
is employed in a number of studies exploiting the role of moisture in ocean-to-land monsoon-like circulations
\citep{Levermann2009,Herzschuh2014,Levermann2016,Boers2017}.
However, as recently pointed out by \citet{Boos2016b}, who quoted \citet{emetal94}, see also \citet{Boos2016a},
the idea that a more intense release of latent heat makes the atmosphere warmer represents {\it "an influential and lengthy dead-end road in atmospheric science"}. Indeed, the steady-state pressure gradients associated with latent heat release
are independent of the intensity of rainfall; they only depend on the steady-state difference
in the amounts of moisture between the rising air and its environment. Again, work/energy and power approaches are confused here.

Thus, a more intense release of latent heat cannot explain the observed correlation between
the circulation intensity and rainfall in either hurricane or monsoon studies.
At this moment the only concept that provides a quantitative explanation to this pattern is
the condensation-induced atmospheric dynamics \citep{pla14,pla15}.
Here key is the positive feedback between the radial air motion and the pressure drop at the surface
associated with condensation and hydrostatic adjustment. As the air streams towards the hurricane
center and ascends, the water vapor condenses and the air pressure drops. The key numerical scale for this
process is the saturated partial pressure of water vapor at the surface (40 hPa at 30 $^{\rm o}$C), which
gives the maximum pressure drop in the {\it power} region of the hurricane.
So far this approach has not won much attention from the meteorological community,
but we hope that it is a matter of the future.

\begin{appendix}
\section*{\begin{center}Appendix A: Deriving Equation (\ref{bud})\end{center}}

\setcounter{equation}{0}%
\renewcommand{\theequation}{A.\arabic{equation}}%

For any quantity $X$ we can write
\beq\label{pro}
\rho (\mathbf{v} \cdot \nabla) X = (\nabla \cdot \mathbf{v}) X \rho  - X \nabla \cdot (\rho \mathbf{v}) =
(\nabla \cdot \mathbf{v}) X \rho  - X \dot{\rho},
\eeq
where in the last equality the steady-state continuity equation (\ref{cont}) has been applied.
Using the definition of material derivative (\ref{der}) and the divergence theorem
we obtain from (\ref{pro})
\beq \label{X1}
\int\limits_{\mathcal{V}}\rho \frac{dX}{dt} d\mathcal{V} = -\int\limits_{\mathcal{V}} \dot{\rho} X  d\mathcal{V}.
\eeq
We have assumed that the circulation is closed such that $\int \mathbf{v}\cdot \mathbf{n} dS = 0$,
where $S$ is the bounding surface of the circulation and $\mathbf{n}$ is the unit normal vector.

Bernoilli equation (\ref{B}) in the form
\beq\label{Bd}
\rho \frac{dK}{dt} = - \frac{dp}{dt} + \rho \mathbf{g} \cdot \mathbf{w} + \mathbf{F} \cdot \mathbf{v},
\eeq
where $K \equiv v^2/2$ is kinetic energy per unit air mass, is obtained from the steady-state equations of motion
\beq\label{mot}
\rho \frac{d \mathbf{v}}{dt} = -\nabla p + \rho \mathbf{g} + \mathbf{F},
\eeq
by taking their scalar product with gas velocity vector $\mathbf{v}$.

Noting that $\mathbf{g} = -g \nabla z$ and using (\ref{X1}) with $X = z$
we obtain for the volume integral of the second term in the right-hand part of (\ref{Bd})
\beq\label{wint}
\int\limits_{\mathcal{V}} \rho \mathbf{g} \cdot \mathbf{w} d\mathcal{V} = -\int\limits_{\mathcal{V}} \rho g \frac{dz}{dt} d\mathcal{V} = \int\limits_{\mathcal{V}} \dot{\rho} g z d\mathcal{V}.
\eeq

Using (\ref{X1}) with $X = K$ and integrating (\ref{Bd}) over $\mathcal{V}$ we find
\beq\label{pbud}
\underbrace{-\int\limits_{\mathcal{V}} \frac{dp}{dt} d\mathcal{V}}_{\rm A} =
\underbrace{- \int\limits_{\mathcal{V}} \mathbf{F} \cdot \mathbf{v} d\mathcal{V}}_{\rm B} \,\,
\underbrace{-\int\limits_{\mathcal{V}} \dot{\rho} g z d\mathcal{V}}_{\rm C}\,\,
\underbrace{-\int\limits_{\mathcal{V}} \dot{\rho} K d\mathcal{V}}_{\rm D}.
\eeq
Work of atmospheric pressure gradients per unit time (A) is balanced by dissipation of kinetic energy (B), production of the gravitational power of precipitation (C)  and production of the kinetic power of gas that is converted to liquid (D).
We have neglected the condensate loading in the equations of motion (\ref{mot}), which implies that all condensate
is instantaneously removed from the atmosphere. Consideration of condensate loading requires a specification of how
it interacts with the air \citep{jgra17}. The resulting term is however small relative to C and can be neglected.

Since typical values of $z$ where condensation occurs are $z \sim 5$~km, we have $gz \gg K$ for any realistic values
of $v$. Even in hurricanes with $v \sim 60$~m~s$^{-1}$, $K$ is only about 4\% of $gz$. Thus term D in (\ref{pbud})
can be neglected without losing accuracy.
\end{appendix}

\section*{Acknowledgment}
This work is par\-ti\-al\-ly supported by  the University of California Agricultural Experiment Station,  Australian Research Council project DP160102107 and the CNPq/CT-Hidro - GeoClima project Grant~404158/2013-7.


\begin{thebibliography}{44}
\expandafter\ifx\csname natexlab\endcsname\relax\def\natexlab#1{#1}\fi
\expandafter\ifx\csname url\endcsname\relax
  \def\url#1{{\tt #1}}\fi
\expandafter\ifx\csname urlprefix\endcsname\relax\def\urlprefix{URL }\fi
\expandafter\ifx\csname doiprefix\endcsname\relax\def\doiprefix{doi:}\fi

\bibitem[{Bister and Emanuel(1998)}]{bister98}
Bister, M. and K.~A. Emanuel, 1998: Dissipative heating and hurricane
  intensity. {\it Meteorol. Atmos. Phys.\/}, {\bf 65}, 233--240,
  doi:10.1007/BF01030791.

\bibitem[{Bister et~al.(2011)Bister, Renno, Pauluis, and Emanuel}]{bister11}
Bister, M., N.~Renno, O.~Pauluis, and K.~Emanuel, 2011: Comment on {Makarieva}
  et al. {\textquoteleft}{A} critique of some modern applications of the
  {Carnot} heat engine concept: the dissipative heat engine cannot
  exist{\textquoteright}. {\it Proceedings of the Royal Society of London A:
  Mathematical, Physical and Engineering Sciences\/}, {\bf 467}, 1--6,
  doi:10.1098/rspa.2010.0087.

\bibitem[{Boers et~al.(2017)Boers, Marwan, Barbosa, and Kurths}]{Boers2017}
Boers, N., N.~Marwan, H.~M.~J. Barbosa, and J.~Kurths, 2017: A
  deforestation-induced tipping point for the {South American} monsoon system.
  {\it Scientific Reports\/}, {\bf 7}, doi:doi:10.1038/srep41489.

\bibitem[{Boos and Storelvmo(2016{\natexlab{a}})}]{Boos2016a}
Boos, W.~R. and T.~Storelvmo, 2016{\natexlab{a}}: Near-linear response of mean
  monsoon strength to a broad range of radiative forcings. {\it Proc. Natl.
  Acad. Sci.\/}, {\bf 113}, 1510--1515, doi:10.1073/pnas.1517143113.

\bibitem[{Boos and Storelvmo(2016{\natexlab{b}})}]{Boos2016b}
--- 2016{\natexlab{b}}: Reply to {Levermann et al.: Linear} scaling for
  monsoons based on well-verified balance between adiabatic cooling and latent
  heat release. {\it Proc. Natl. Acad. Sci.\/}, {\bf 113}, E2350--E2351,
  doi:10.1073/pnas.1603626113.

\bibitem[{Camp and Montgomery(2001)}]{CampMontgomery01}
Camp, J.~P. and M.~T. Montgomery, 2001: Hurricane maximum intensity: {Past} and
  present. {\it Mon. Weather Rev.\/}, {\bf 129}, 1704--1717,
  doi:10.1175/1520-0493(2001)129<1704:HMIPAP>2.0.CO;2.

\bibitem[{Curzon and Ahlborn(1975)}]{curzon75}
Curzon, F.~L. and B.~Ahlborn, 1975: Efficiency of a {Carnot} engine at maximum
  power output. {\it Am. J. Phys.\/}, {\bf 43}, 22--24, doi:10.1119/1.10023.

\bibitem[{DeMaria and Kaplan(1994)}]{demaria94}
DeMaria, M. and J.~Kaplan, 1994: Sea surface temperature and the maximum
  intensity of {Atlantic} tropical cyclones. {\it J. Climate\/}, {\bf 7},
  1324--1334, doi:10.1175/1520-0442(1994)007<1324:SSTATM>2.0.CO;2.

\bibitem[{Emanuel(1986)}]{em86}
Emanuel, K.~A., 1986: An air-sea interaction theory for tropical cyclones.
  {Part I: Steady-state} maintenance. {\it J. Atmos. Sci.\/}, {\bf 43},
  585--604, doi:10.1175/1520-0469(1986)043<0585:AASITF>2.0.CO;2.

\bibitem[{Emanuel(1988)}]{em88}
--- 1988: The maximum intensity of hurricanes. {\it J. Atmos. Sci\/}, {\bf 45},
  1143--1155, doi:10.1175/1520-0469(1988)045<1143:TMIOH>2.0.CO;2.

\bibitem[{Emanuel(1991)}]{em91}
--- 1991: The theory of hurricanes. {\it Annu. Rev. Fluid Mech.\/}, {\bf 23},
  179--196, doi:10.1146/annurev.fl.23.010191.001143.

\bibitem[{Emanuel(1995)}]{em95}
--- 1995: Sensitivity of tropical cyclones to surface exchange coefficients and
  a revised steady-state model incorporating eye dynamics. {\it J. Atmos.
  Sci\/}, {\bf 52}, 3969--3976,
  doi:10.1175/1520-0469(1995)052<3969:SOTCTS>2.0.CO;2.

\bibitem[{Emanuel(1997)}]{em97}
--- 1997: Some aspects of hurricane inner-core dynamics and energetics. {\it J.
  Atmos. Sci\/}, {\bf 54}, 1014--1026,
  doi:10.1175/1520-0469(1997)054<1014:SAOHIC>2.0.CO;2.

\bibitem[{Emanuel et~al.(1994)Emanuel, David~Neelin, and Bretherton}]{emetal94}
Emanuel, K.~A., J.~David~Neelin, and C.~S. Bretherton, 1994: On large-scale
  circulations in convecting atmospheres. {\it Q. J. R. Meteorol. Soc.\/}, {\bf
  120}, 1111--1143, doi:10.1002/qj.49712051902.

\bibitem[{Ermakov et~al.(2014)Ermakov, Sharkov, and Chernushich}]{ermakov14}
Ermakov, D.~M., E.~A. Sharkov, and A.~P. Chernushich, 2014: The role of
  tropospheric advection of latent heat in the intensification of tropical
  cyclones. {\it Issledovanie Zemli iz Kosmosa, (Earth Research from Space)\/},
  {\bf 4}, 3--15, doi:10.7868/S0205961414040046, (in Russian).

\bibitem[{Ermakov et~al.(2015)Ermakov, Sharkov, and Chernushich}]{ermakov15}
--- 2015: Satellite radiothermovision of atmospheric mesoscale processes: case
  study of tropical cyclones. {\it Int. Arch. Photogramm. Remote Sens. Spatial
  Inf. Sci.\/}, {\bf XL-7/W3}, 179--186,
  doi:10.5194/isprsarchives-XL-7-W3-179-2015.

\bibitem[{Frank(1977)}]{frank77}
Frank, W.~M., 1977: The structure and energetics of the tropical cyclone {I.
  Storm} structure. {\it Mon. Weather Rev.\/}, {\bf 105}, 1119--1135,
  doi:10.1175/1520-0493(1977)105<1119:TSAEOT>2.0.CO;2.

\bibitem[{Fritz and Wang(2014)}]{fritz14}
Fritz, C. and Z.~Wang, 2014: Water vapor budget in a developing tropical
  cyclone and its implication for tropical cyclone formation. {\it J. Atmos.
  Sci.\/}, {\bf 71}, 4321--4332, doi:10.1175/JAS-D-13-0378.1.

\bibitem[{Fujiwara et~al.(2017)Fujiwara, Kawamura, Hirata, Kawano, Kato, and
  Shinoda}]{fujiwara17}
Fujiwara, K., R.~Kawamura, H.~Hirata, T.~Kawano, M.~Kato, and T.~Shinoda, 2017:
  A positive feedback process between tropical cyclone intensity and the
  moisture conveyor belt assessed with lagrangian diagnostics. {\it J. Geophys.
  Res. Atmos.\/}, {\bf 122}, 12502--12521, doi:10.1002/2017JD027557.

\bibitem[{Gorshkov and Dol'nik(1980)}]{vg80}
Gorshkov, V.~G. and V.~R. Dol'nik, 1980: Energetics of the biosphere. {\it
  Soviet Physics Uspekhi\/}, {\bf 23}, 386--408,
  doi:10.1070/PU1980v023n07ABEH005117.

\bibitem[{Herzschuh et~al.(2014)Herzschuh, Borkowski, Schewe, Mischke, and
  Tian}]{Herzschuh2014}
Herzschuh, U., J.~Borkowski, J.~Schewe, S.~Mischke, and F.~Tian, 2014:
  Moisture-advection feedback supports strong early-to-mid {Holocene} monsoon
  climate on the eastern {Tibetan Plateau} as inferred from a pollen-based
  reconstruction. {\it Palaeogeography, Palaeoclimatology, Palaeoecology\/},
  {\bf 402}, 44 -- 54, doi:https://doi.org/10.1016/j.palaeo.2014.02.022.

\bibitem[{Holland(1997)}]{holland97}
Holland, G.~J., 1997: The maximum potential intensity of tropical cyclones.
  {\it J. Atmos. Sci.\/}, {\bf 54}, 2519--2541,
  doi:10.1175/1520-0469(1997)054<2519:TMPIOT>2.0.CO;2.

\bibitem[{Jaimes et~al.(2015)Jaimes, Shay, and Uhlhorn}]{jaimes15}
Jaimes, B., L.~K. Shay, and E.~W. Uhlhorn, 2015: Enthalpy and momentum fluxes
  during {Hurricane Earl} relative to underlying ocean features. {\it Mon.
  Weather Rev.\/}, {\bf 143}, 111--131, doi:10.1175/MWR-D-13-00277.1.

\bibitem[{Kieu(2015)}]{kieu15}
Kieu, C., 2015: Revisiting dissipative heating in tropical cyclone maximum
  potential intensity. {\it Quart. J. Roy. Meteorol. Soc.\/}, {\bf 141},
  2497--2504, doi:10.1002/qj.2534.

\bibitem[{Krishnamurti et~al.(1993)Krishnamurti, Bedi, and
  Ingles}]{krishnamurti93}
Krishnamurti, T.~N., H.~S. Bedi, and K.~Ingles, 1993: Physical initialization
  using {SSM/I} rain rates. {\it Tellus A\/}, {\bf 45}, 247--269,
  doi:10.1034/j.1600-0870.1993.t01-3-00001.x.

\bibitem[{Krishnamurti et~al.(1998)Krishnamurti, Han, and
  Oosterhof}]{krishnamurti98}
Krishnamurti, T.~N., W.~Han, and D.~Oosterhof, 1998: Sensitivity of hurricane
  intensity forecasts to physical initialization. {\it Meteorol. Atmos.
  Phys.\/}, {\bf 65}, 171--181, doi:10.1007/BF01030786.

\bibitem[{Levermann et~al.(2016)Levermann, Petoukhov, Schewe, and
  Schellnhuber}]{Levermann2016}
Levermann, A., V.~Petoukhov, J.~Schewe, and H.~J. Schellnhuber, 2016: Abrupt
  monsoon transitions as seen in paleorecords can be explained by
  moisture-advection feedback. {\it Proc. Natl. Acad. Sci.\/}, {\bf 113},
  E2348--E2349, doi:10.1073/pnas.1603130113.

\bibitem[{Levermann et~al.(2009)Levermann, Schewe, Petoukhov, and
  Held}]{Levermann2009}
Levermann, A., J.~Schewe, V.~Petoukhov, and H.~Held, 2009: Basic mechanism for
  abrupt monsoon transitions. {\it Proc. Natl. Acad. Sci.\/}, {\bf 106},
  20572--20577, doi:10.1073/pnas.0901414106.

\bibitem[{Makarieva et~al.(2010)Makarieva, Gorshkov, Li, and Nobre}]{dhe10}
Makarieva, A.~M., V.~G. Gorshkov, B.-L. Li, and A.~D. Nobre, 2010: A critique
  of some modern applications of the {Carnot} heat engine concept: the
  dissipative heat engine cannot exist. {\it Proc. R. Soc. A\/}, {\bf 466},
  1893--1902, doi:10.1098/rspa.2009.0581.

\bibitem[{Makarieva et~al.(2014)Makarieva, Gorshkov, and Nefiodov}]{pla14}
Makarieva, A.~M., V.~G. Gorshkov, and A.~V. Nefiodov, 2014: Condensational
  power of air circulation in the presence of a horizontal temperature
  gradient. {\it Phys. Lett. A\/}, {\bf 378}, 294--298,
  doi:10.1016/j.physleta.2013.11.019.

\bibitem[{Makarieva et~al.(2015)Makarieva, Gorshkov, and Nefiodov}]{pla15}
--- 2015: Empirical evidence for the condensational theory of hurricanes. {\it
  Phys. Lett. A\/}, {\bf 379}, 2396--2398, doi:10.1016/j.physleta.2015.07.042.

\bibitem[{Makarieva et~al.(2017{\natexlab{a}})Makarieva, Gorshkov, Nefiodov,
  Chikunov, Sheil, Nobre, and Li}]{ar17}
Makarieva, A.~M., V.~G. Gorshkov, A.~V. Nefiodov, A.~V. Chikunov, D.~Sheil,
  A.~D. Nobre, and B.-L. Li, 2017{\natexlab{a}}: Fuel for cyclones: The water
  vapor budget of a hurricane as dependent on its movement. {\it Atmospheric
  Research\/}, {\bf 193}, 216--230, doi:10.1016/j.atmosres.2017.04.006.

\bibitem[{Makarieva et~al.(2013)Makarieva, Gorshkov, Nefiodov, Sheil, Nobre,
  Bunyard, and Li}]{jas13}
Makarieva, A.~M., V.~G. Gorshkov, A.~V. Nefiodov, D.~Sheil, A.~D. Nobre,
  P.~Bunyard, and B.-L. Li, 2013: The key physical parameters governing
  frictional dissipation in a precipitating atmosphere. {\it J. Atmos. Sci.\/},
  {\bf 70}, 2916--2929, doi:10.1175/JAS-D-12-0231.1.

\bibitem[{Makarieva et~al.(2017{\natexlab{b}})Makarieva, Gorshkov, Nefiodov,
  Sheil, Nobre, Bunyard, Nobre, and Li}]{jgra17}
Makarieva, A.~M., V.~G. Gorshkov, A.~V. Nefiodov, D.~Sheil, A.~D. Nobre,
  P.~Bunyard, P.~Nobre, and B.-L. Li, 2017{\natexlab{b}}: The equations of
  motion for moist atmospheric air. {\it J. Geophys. Res. Atmos.\/}, {\bf 122},
  7300--7307, doi:10.1002/2017JD026773.

\bibitem[{Makarieva et~al.(2017{\natexlab{c}})Makarieva, Gorshkov, Nefiodov,
  Sheil, Nobre, Shearman, and Li}]{tellus17}
Makarieva, A.~M., V.~G. Gorshkov, A.~V. Nefiodov, D.~Sheil, A.~D. Nobre, P.~L.
  Shearman, and B.-L. Li, 2017{\natexlab{c}}: Kinetic energy generation in heat
  engines and heat pumps: The relationship between surface pressure,
  temperature and circulation cell size. {\it Tellus A\/}, {\bf 69}, 1272752,
  doi:10.1080/16000870.2016.1272752.

\bibitem[{{Makarieva et al.}(in preparation)}]{prep18}
{Makarieva et al.}, in preparation: {Hurricane's Maximum Potential Intensity},
  hypercanes and fluxes of sensible and latent heat.

\bibitem[{Malkus and Riehl(1960)}]{malkus60}
Malkus, J.~S. and H.~Riehl, 1960: On the dynamics and energy transformations in
  steady-state hurricanes. {\it Tellus\/}, {\bf 12}, 1--20,
  doi:10.1111/j.2153-3490.1960.tb01279.x.

\bibitem[{Pauluis(2011)}]{pa11}
Pauluis, O., 2011: Water vapor and mechanical work: {A} comparison of {Carnot}
  and steam cycles. {\it J. Atmos. Sci.\/}, {\bf 68}, 91--102,
  doi:10.1175/2010JAS3530.1.

\bibitem[{Pauluis et~al.(2000)Pauluis, Balaji, and Held}]{pa00}
Pauluis, O., V.~Balaji, and I.~M. Held, 2000: Frictional dissipation in a
  precipitating atmosphere. {\it J. Atmos. Sci.\/}, {\bf 57}, 989--994,
  doi:10.1175/1520-0469(2000)057<0989:FDIAPA>2.0.CO;2.

\bibitem[{Pauluis(2015)}]{pa15}
Pauluis, O.~M., 2015: The global engine that could. {\it Science\/}, {\bf 347},
  475--476, doi:10.1126/science.aaa3681.

\bibitem[{Pauluis and Zhang(2017)}]{pa17}
Pauluis, O.~M. and F.~Zhang, 2017: Reconstruction of thermodynamic cycles in a
  high-resolution simulation of a hurricane. {\it Journal of the Atmospheric
  Sciences\/}, {\bf 74}, 3367--3381, doi:10.1175/JAS-D-16-0353.1.

\bibitem[{Sabuwala et~al.(2015)Sabuwala, Gioia, and Chakraborty}]{sabuwala15}
Sabuwala, T., G.~Gioia, and P.~Chakraborty, 2015: Effect of rainpower on
  hurricane intensity. {\it Geophys. Res. Lett.\/}, {\bf 42}, 3024--3029,
  doi:10.1002/2015GL063785.

\bibitem[{Smith et~al.(2014)Smith, Montgomery, and Persing}]{smith14}
Smith, R.~K., M.~T. Montgomery, and J.~Persing, 2014: On steady-state tropical
  cyclones. {\it Quart. J. Roy. Meteorol. Soc.\/}, {\bf 140}, 2638--2649,
  doi:10.1002/qj.2329.

\bibitem[{Smith et~al.(2008)Smith, Montgomery, and Vogl}]{sm08}
Smith, R.~K., M.~T. Montgomery, and S.~Vogl, 2008: A critique of {Emanuel's}
  hurricane model and potential intensity theory. {\it Q. J. R. Meteorol.
  Soc.\/}, {\bf 134}, 551--561.

\end{thebibliography}

\clearpage

\end{document}